\documentclass[letterpaper,prx,aps,twocolumn,showpacs,superscriptaddress,nofootinbib]{revtex4-2}
\usepackage{bm,color}
\usepackage[T1]{fontenc}
\usepackage[utf8]{inputenc}
\usepackage{latexsym}
\usepackage{graphicx} 
\usepackage{amsmath} 
\usepackage{amsfonts} 
\usepackage{braket}
\usepackage{subfigure}
\usepackage{bbold}
\usepackage{lineno}
\usepackage{hyperref}

\newcommand{\bbm}{\begin{pmatrix}}
\newcommand{\ebm}{\end{pmatrix}}	

\newcommand{\bk}{\mathbf{k}}

\usepackage[normalem]{ulem} 
\definecolor{mgreen}{RGB}{10,160,20}

\def\be{\begin{equation}}
\def\ee{\end{equation}}

\begin{document}

\title{Micromotion area as proxy for anomalous Floquet topological systems}

\author{Luca Asteria}
\email{asteria.luca.6w@kyoto-u.ac.jp}
\affiliation{Department of Physics, Graduate School of Science, Kyoto University, Kyoto 606-8502, Japan}

\author{Klaus Sengstock}
\affiliation{Institute for Quantum Physics, University of Hamburg, 22761 Hamburg, Germany}
\affiliation{The Hamburg Centre for Ultrafast Imaging, 22761 Hamburg, Germany}

\author{André Eckardt}
\affiliation{Institut für Physik und Astronomie, Technische Universität Berlin, 10623 Berlin, Germany}

\author{Christof Weitenberg}
\email {christof.weitenberg@tu-dortmund.de}
\affiliation{Department of Physics, TU Dortmund University, 44227 Dortmund, Germany}


\begin{abstract}
Driven Floquet systems can realize topological phases with no static counterparts. These so-called anomalous Floquet topology breaks the bulk-boundary correspondence based on the Chern number. The number of edge modes in each band gap is instead determined by another integer index, a winding number, which is calculated from the time evolution operator of the bulk states within one driving period. While in the non-driven system, Chern markers provide a useful local proxy for the Chern number in the bulk, so far no such local bulk indicator is known for the winding number in Floquet systems.
Here we consider two-band models and show that the area enclosed during a Floquet period by an initially localized particle signals the presence of an anomalous phase when it approaches half the unit cell area. In general, we show that at the fine-tuned point of dispersionless dynamics during the micromotion, the enclosed area is quantized and an exact proportionality relation exists between the area and the winding number. Direct detection of anomalous topology in real space could be realized in several quantum simulation platforms, and could be useful for systems with disorder or interactions. Building on the connection between area and winding number, we also show a way to realize arbitrarily high winding numbers.
\end{abstract}

\maketitle

The description by Landau for phase transitions based on a local order parameter does not apply to topological states, where order is not local but encoded in topological quantum numbers \cite{Hasan2010}. Topological states can be classified according to dimensionality and symmetry of the system \cite{Schnyder2008_classification,kitaev2009,Ryu_2010}, to which different topological indices correspond and need to be used to describe the system and its properties. A paradigmatic example is the Chern number, describing a quantized Hall response as well as the number of chiral edge modes (bulk-edge correspondence) in two dimensional systems.

The relevant topological description might, however, change when the system is periodically driven and described via Floquet theory \cite{eckardtColloquiumFloquet,Weitenberg2021}. In particular, in the case where the driving period $T$ becomes long enough to approach the time scale of the unmodulated system a different topological classification is needed: As quasi-energy is defined only up to an integer multiples of $2\pi$ (in units of $\hbar/T$), chiral edge modes can exist also in the so-called $\pi$ band gap, which separates quasienergy bands centered around $0$ with equivalent ones centered around $2\pi$ in an extended zone scheme. This allows for anomalous Floquet topological states, where chiral edge modes exist, while at the same time all quasienergy bands have a trivial Chern number of zero \cite{Kitagawa2010,Rudner2013,GomezLeon2024anomalousfloquet}. A complete classification is then given by a set of integer winding numbers $W_g$, associated with the band gaps (labeled by $g$) in the system. Such anomalous Floquet topological phases have been realized in photonics \cite{Maczewsky2017,Mukherjee2017}, acoustic systems \cite{Peng2016}, and cold atoms \cite{Wintersperger2020}. 

In the above mentioned experiments, the topologically non-trivial properties of the system were either probed via chiral edge motion \cite{Peng2016,Maczewsky2017,Braun2024}, invoking the bulk-boundary correspondence, or by using momentum-resolved measurements of bulk properties, which are non-local in real space \cite{Uenal2019,Wintersperger2020}. While Chern insulators have a clear real space bulk signature as well, as revealed through the quantization of the transverse Hall response \cite{Aidelsburger2015} and local Chern markers \cite{Chalopin2020,BiancoResta2011}, such a {local bulk} signature was not yet identified so far for anomalous topological systems. 

\begin{figure}[h!]	\includegraphics[width=0.85\linewidth]{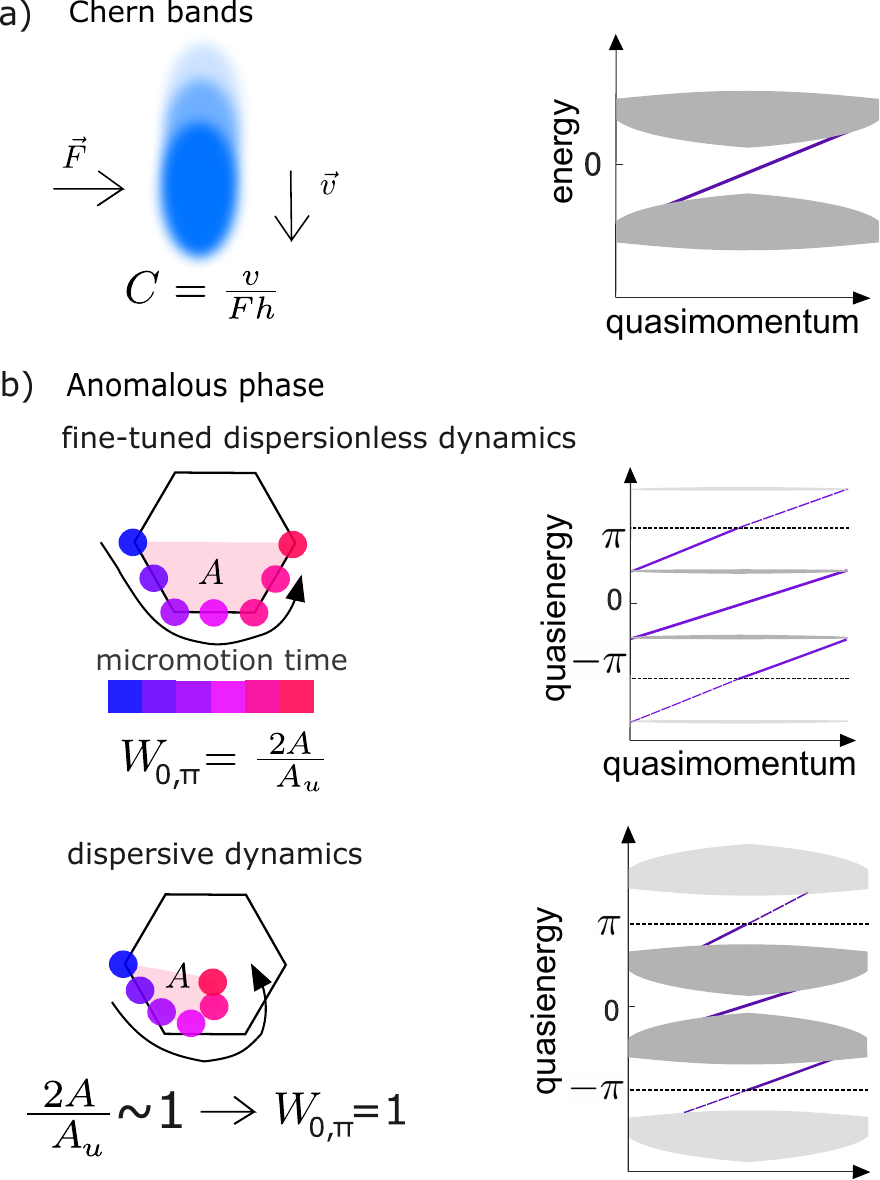}
\caption{{Bulk observables for Chern systems and anomalous systems} \small a) In static systems, the Chern number $C$ determines the number of chiral edge modes (right panel) and can be detected in the bulk by the quantized transverse conductance (left panel).  b) In anomalous Floquet systems, the edge modes are given by the winding numbers $W_g$ for each band gap $g$. We show that - in the case of dispersionless {micromotion dynamics} - the micromotion area $A$ is also quantized and indicates the winding numbers while in the more general case, a finite micromotion area can be used to infer the presence of the anomalous phase.}
    \label{fig:1}
\end{figure}

In the following, we propose a simple local and experimentally accessible indicator also for anomalous Floquet topological systems, given by the area encircled by an initially localized particle. We study the link between the winding number and the real-space dynamics in the bulk of the system for generic lattice models with two sublattice states $s\in\{0,1\}$ and cyclic tunnel modulation. In particular, we consider the area $A$ encircled during one driving cycle by the center of mass of particles initially localized on a single lattice site of either sublattice (as the area does not depend on the sublattice). We refer to $A$ as micromotion area, as it predominantly results from the Floquet micromotion.
We find that if the micromotion dynamics is completely dispersionless (fine-tuned point), the system is in an anomalous phase with equal winding numbers in both gaps and that these are directly given by twice the encircled area $A$ in units of the unit area $A_u$, $W_0=W_\pi=2A/A_u$ (see End Matter~\ref{supp:fine-tuned}). Away from the fine-tuned point, the area is no longer quantized. However, our numerical results show that values of $\frac{2A}{A_u}$ of the order of one, still provide a proxy that allows to identify the anomalous regime as well as its breakdown. 
Finally, we show that the relation between micromotion area and winding number can be used to devise protocols for anomalous phases with arbitrarily high winding numbers or, equivalently, detection of the area can be used to probe such systems (Fig.~\ref{fig:1}). 

These results are complementary to recent work relating the Streda formula to the winding number of Floquet topological systems \cite{gavensky2024}). 
We note that also in 1D systems, topology can be extracted from real-space dynamics - in that case in the long-time limit - via the mean chiral displacement as a measure for the Zak phase \cite{Cardano2017}, which also applies to Floquet systems \cite{Maffei2018}.

{\it Relation between $W$ and $A$ - }
\label{adiabatic_Winding_Number}
The winding number, which counts the number of edge modes in a gap $g$ is defined as \cite{Rudner2013}:
\begin{equation}
\begin{aligned}
    W_{g}=&\frac{1}{8\pi^2}\int_0^{2T} \mathrm{d}t \int_{BZ}\mathrm{d}{\mathbf{k}}~\mathrm{Tr}~\hat{O}\\
    \hat{O}=&-{U}_g^{-1}\partial_t {U}_g~ \Bigl[~{U}_g^{-1}\partial_{k_x} {U}_g, {U}_g^{-1}\partial_{k_y} {U}_g~\Bigr],
    \end{aligned}
    \label{WindingN_definition}
\end{equation}
where, here and in the following, the partial derivatives are assumed to act only on the directly neighboring term to their right and where
${U}_g$ is the band-flattend time-evolution operator (in this work, a $2\times2$ matrix), which depends on the wavevector (in the first Brillouin zone, denoted by $BZ$). 

It is defined as
\begin{equation}
 {U}_g(\bk,t)=
\begin{cases}
U(\bk,t),&0\le t\le T\\
\exp\big(-\frac{i}{\hbar}(2T-t)H_f^g(\bk)\big), & T<t\le 2T,
\end{cases}
 \end{equation}
where $U(\mathbf{k},t)$ denotes the physical time evolution operator and where 
\begin{equation}
  {\hat{H}^g_f(k)=\frac{i\hbar}{T}\log\bigl[{U}(\bk,t=T)~\bigr]_\text{branch cut at $g/T$}}
\end{equation} 
is the Floquet Hamiltonian, with the branch cut of the logarithm chosen to lie in the band gap $g$.

\begin{figure}[ht]
\includegraphics[width=0.9\linewidth]{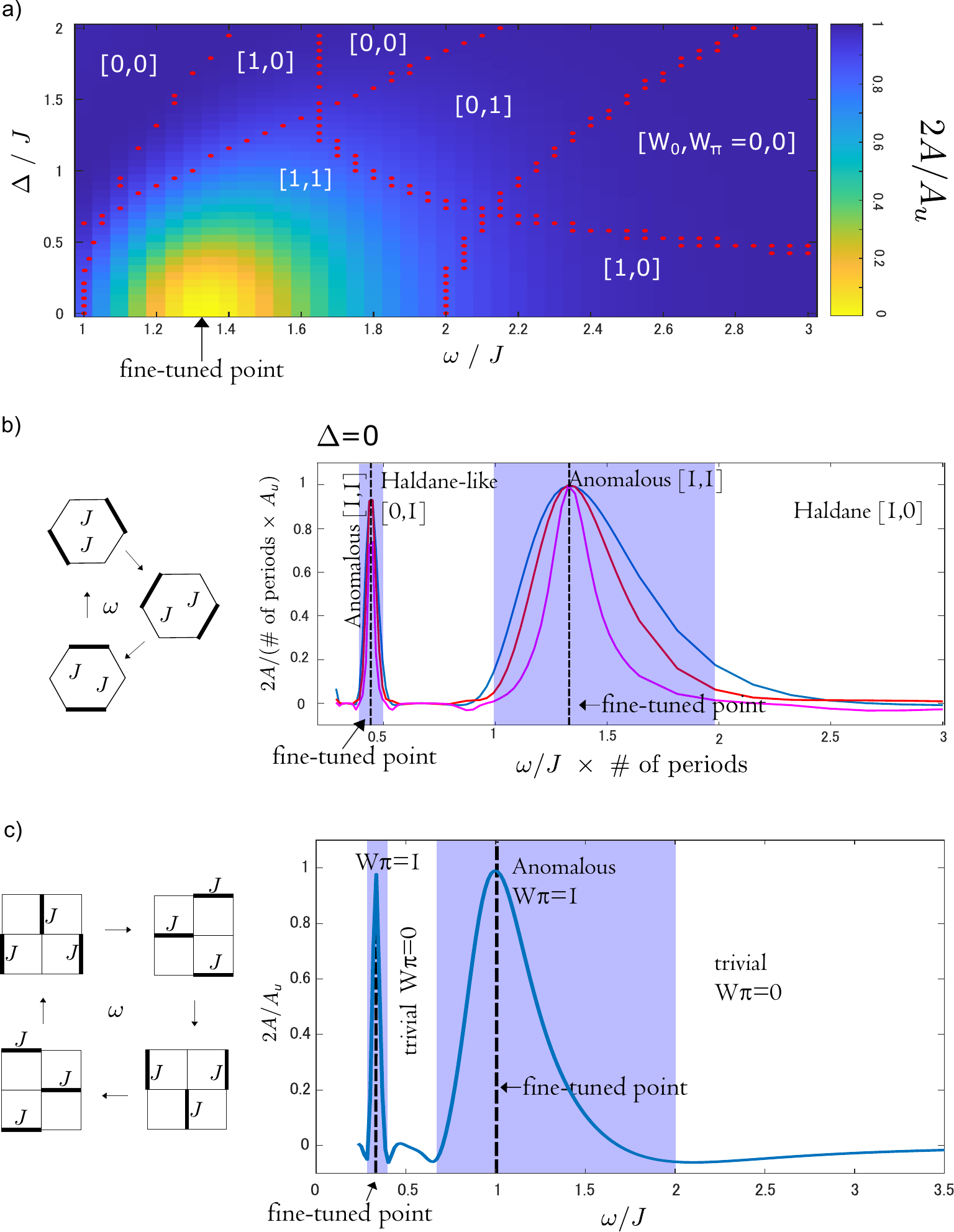}
\caption{{Micromotion area as a function of $\omega$ and $\Delta$}.
{\bf a}) Tunneling in a honeycomb lattice is completely switched on and off in a cyclic way  $J_1\xrightarrow{}J_2\xrightarrow{}J_3\xrightarrow{}...$ with frequency $\omega$. A constant sublattice offset $\Delta$ is present.
We plot the area calculated across the phase diagram. Red dots mark topological phase transition points, and the brackets indicate $[W_0,W_\pi]$ (calculated in \cite{Martinez2023}). {\bf b)} Area (blue curve) as a function of the Floquet frequency for $\Delta=0$. An additional anomalous phase appears for even lower frequency drives \cite{Wintersperger2020}. The anomalous phases are indicated by the purple shade. The red (magenta) curve shows the area for two (five) Floquet periods. {\bf c})  Area as a function of the Floquet frequency (right) in the cyclic protocol (left) in the bipartite square lattice as introduced in \cite{Rudner2013}. The $0$ gap is always closed for these protocols, and only the value of $W_\pi$ is noted. Also here anomalous phases for smaller driving frequencies are present (purple shades).} 
\label{fig:2}
\end{figure}

In order to establish the connection between the encircled area $A$ and the winding number, we consider the initial state 
\begin{equation}
\ket{\psi_0}=\sqrt{\frac{A_u}{4\pi^2}}\int_{BZ}\mathrm{d}{\mathbf{k}}~\ket{0,\mathbf{k}}
\end{equation}
that is localized on a single site of sublattice $s=0$ at the origin.
Here $\ket{s,\mathbf{k}}$ denotes the Bloch state for sublattice $s$ with wavevector $\mathbf{k}$.
The mean position of the time-evolved state is given by \cite{Vanderbilt_2018}:
\begin{equation}
    {\bf r}(t)=i\frac{A_u}{4\pi^2}\int_{BZ}\mathrm{d}{\bf k}~\bra{0}U^{-1}\nabla{\bf }~U\ket{0}, 
\end{equation}
whereas the velocity is given by
\begin{equation}
    \begin{aligned}
        {\bf v}=\partial_t {\bf r}=&\frac{A_u}{4\pi^2}\int_{BZ}\mathrm{d}{\bf k}\frac{1}{i\hbar}\bra{0}U^{-1}[~i\nabla,H(t)~]~U\ket{0}\\
        =&\frac{A_u}{4\pi^2\hbar}\int_{BZ}\mathrm{d}{\bf k}~~\bra{0}U^{-1}~(\nabla H(t))~U\ket{0}\\
        =&~~~~\frac{A_u}{4\pi^2\hbar}\int_{BZ}\mathrm{d}{\bf k}~\bra{0}\nabla (U^{-1}HU)\ket{0}\\
        &-\frac{A_u}{4\pi^2\hbar}\int_{BZ}\mathrm{d}{\bf k}~\bra{0}\nabla U^{-1}HU\ket{0}\\
        &-\frac{A_u}{4\pi^2\hbar}\int_{BZ}\mathrm{d}{\bf k}~\bra{0}U^{-1}H\nabla U\ket{0}\\=&~~~~\frac{A_u}{4\pi^2\hbar}\int_{BZ}\mathrm{d}{\bf k}~\bra{0}\nabla (U^{-1}HU)\ket{0}\\
        &+\frac{A_u}{4\pi^2i}\int_{BZ}\mathrm{d}{\bf k}~\bra{0}\nabla U^{-1}\partial_t U\ket{0}\\
        &-\frac{A_u}{4\pi^2i}\int_{BZ}\mathrm{d}{\bf k}~\bra{0}\partial_tU^{-1}\nabla U\ket{0}\\
    \end{aligned}
\end{equation}
$U^{-1}H=\frac{\hbar}{i}\partial_tU^{-1}$ and $HU={i\hbar}\partial_tU$.
Note that the first term can be regarded as the normal component of the velocity, while the remaining two terms correspond to the anomalous velocity as given by the Berry curvature $\Omega_\nabla=\bra{0}\nabla U^{-1}\partial_tU-\partial_t U\nabla U\ket{0}$. 
The normal velocity term averages to zero when integrated over the $BZ$. The remaining terms are complex conjugates of each other since $\nabla U^{-1}\partial_tU$ and $\partial_tU^{-1}\nabla U$ are adjoint to each other and therefore we get:
\begin{equation}
    \begin{aligned}
        {\bf v}(t)=&-2\frac{A_u}{4\pi^2}\int_{BZ}\mathrm{d}{\bf k}~\mathrm{Im}[\bra{0}\partial_tU^{-1}\nabla U\ket{0}]\\
    \end{aligned}\label{eq:velocity}
\end{equation}
From ${\bf r}$ and ${\mathbf{v}}$ we can immediately define the area.

\begin{equation}
    A=-\frac{1}{2}\int_0^T \mathrm{d}t~ {\bf v}(t)\times{\bf r}(t)\ \label{eq:area}
\end{equation}
where $\times$ denotes the cross product.
If ${\bf r}$ does not describe a closed trajectory (${\bf r}(0)\neq{\bf r}(T)$) the area so calculated corresponds to the closed area obtained by connecting linearly ${\bf r}(T)$ with ${\bf r}(0)$.
We note that the area in Eq.~(\ref{eq:area}) does not depend on the starting sublattice, because both ${\bf v}(t)$ and ${\bf r}(t)$ change their sign when switching the sublattice. It holds for the velocity ${\bf v}(t)$ as the Berry curvature $\Omega_\nabla$ is opposite for the two sublattices of a two-band model. It also holds for the displacement ${\bf r}(t)$, because it is given by an integral over the velocity.
Using our expressions for ${\bf v}$ and ${\bf r}$, one obtains
\begin{equation}
\begin{aligned}
    A&=-\int_0^T \mathrm{d}t~\frac{A_\mathrm{u}}{4\pi^2}\int_{BZ}\mathrm{d}{\bf k}~ \mathrm{Im}\bigl[\bra{0}\partial_t {U}^{-1}\partial_{k_x} {U}\ket{0}\bigr] \\ &\times i\frac{A_\mathrm{u}}{4\pi^2}\int_{BZ}\mathrm{d}{\mathbf{k}}~\bra{0}{U}^{-1}\partial_{k_y} {U}\ket{0}-x\Longleftrightarrow y
    \end{aligned}
    \label{Area_with_operators}
\end{equation}
{On the other hand, rewriting Eq.~(\ref{WindingN_definition}) for the winding number with the help of $-{U}_g^{-1}\partial_t {U}_g{U}_g^{-1}=\partial_t {U}_g^{-1}$ gives}
\begin{equation}
\begin{aligned}
    W_{g}=&\frac{1}{8\pi^2}\int_0^{2T} \mathrm{d}t \int_{BZ}\mathrm{d}{\mathbf{k}}~\mathrm{Tr}\Bigl(\partial_t {U}_g^{-1}~ \partial_{k_x} {U}_g~ {U}_g^{-1}\partial_{k_y} {U}_g~\Bigr)\\&-k_x\Longleftrightarrow k_y.
    \end{aligned}
    \label{WindingN_definition_2}
\end{equation}
which bears similarity to the expression for the area. The complete calculation (see End Matter~\ref{Full_Calculation})  then leads to
\begin{equation}\begin{aligned}
    W_g=~&\frac{2A}{A_u}+\frac{1}{8\pi^2}\int_T^{2T} \mathrm{d}t \int_{BZ}\mathrm{d}{\mathbf{k}}~\mathrm{Re~(Tr}\Bigl[B\times C\Bigr])\\
    &+\frac{1}{A_u}\sum_{s=0}^1\int_0^{T}\mathrm{d}t~\overline{\mathrm{Im}\bigl[\bra{s}\delta B\ket{s}\bigr]\times\bra{s}\delta C\ket{s}},
    \end{aligned}
    \label{main_equation}
\end{equation}
where $B\equiv\partial_t {U}_g^{-1}\nabla_\bk {U}_g$, $C\equiv {U}_g^{-1}\nabla_\bk {U}_g$,  $\delta B=B-\overline{B},$ and $~\delta C=C-\overline{C}$. The overbar indicates average over the BZ $\overline{f}=\frac{A_u}{4\pi^2}\int_{BZ}\mathrm{d}{\bf k}~f({\bf k})$.
The trace in Eq.~(\ref{WindingN_definition_2}) involves a sum over both sublattices, which amounts to the factor of two on the rhs of Eq.~(\ref{main_equation}), as both sublattices give rise to the same area $A$.
We refer to $\mathrm{Tr}[B\times C]$ as the "band-flattening term" and to $\delta B\times \delta C$ as the "position-velocity correlation term". Note that the band-flattening term, just like $W_g$ depends on the choice of the branch cut, while the other terms do not. Also note that while at the fine-tuned point, flat physical bands are realized, these are not to be confused with the mathematical band-flattening procedure necessary for the winding number to be well defined \cite{Rudner2013}.
Both the band-flattening and the correlation term tend to zero at the fine-tuned point of dispersionless micromotion dynamics (see End Matter~\ref{supp:fine-tuned}) and then a perfect quantization is obtained:
\begin{equation}
    W_0=W_\pi=\frac{2A}{A_u}.
\end{equation}
Note that $W_0=W_\pi$ implies that the Chern number of both bands is zero. Therefore, in these phases, the area is a direct proxy for anomalous topology {(in the sense of $W_0=W_\pi$)} {and it can distinguish it from Haldane-like phases with $W_0\neq W_\pi$ and finite Chern number (Fig.~\ref{fig:2}a).}

{\it Simulations - } Using the example of a hexagonal lattice, we will now investigate the link between the micromotion area and the winding number numerically. In particular, we will also investigate the situation away from the fine-tuned point, where no one-to-one correspondence is expected. However, we will see that the micromotion area still provides a proxy that allows to identify the Floquet anomalous regime.  
In Fig.~\ref{fig:2} we study the behavior of the area as a function of the Floquet frequency $\omega$ and of a constant sublattice offset $\Delta$ for a step-drive protocol~\cite{Kitagawa2010}
\begin{equation}
\begin{aligned}
    J_1&=J,~J_{2,3}=0~~~~~~~0<t<T/3\\
    J_2&=J,~J_{1,3}=0~~~~~~~T/3<t<2T/3\\
    J_3&=J,~J_{1,2}=0~~~~~~~2T/3<t<T\\
\end{aligned}
\end{equation}
with $J_i$ being the time-dependent tunneling matrix elements along the three directions $i$ (Fig.~\ref{fig:2}a) and tunneling strength $J$. The fine-tuned point is realized for $\omega=\frac{3}{4}J$ \cite{Kitagawa2010,Quelle2017}. 

For the high-frequency limit $\omega/J\gg1$ the system can be mapped to the Haldane model where a phase transition from a Chern to a trivial system happens for large sublattice offset $\Delta$. Note that in contrast to related work \cite{Rudner2013}, $\Delta$ is applied during the entire protocol. We obtain different phases as a function of the Floquet frequency and sublattice offset characterized by different values of the winding numbers in the two gaps (compare with Ref.~\cite{Martinez2023}), but $2A/A_u$ approaches unity only in the anomalous phase, characterized by $W_0=W_\pi=1$.

At the fine-tuned point, the area approaches the winding number and numerically we observe that it also rather clearly indicates the whole anomalous regime by assuming values of the order one only here. We note that the area quickly drops to zero when entering the Haldane phase with $[W_0,W_\pi]=[1,0]$ or the Haldane-like phase with $[W_0,W_\pi]=[0,1]$ \cite{Wintersperger2020} (for not too large sublattice offsets), where although the Floquet frequency is still comparable with the system's frequency scales, the system is described in good approximation by an effective Hamiltonian under which no significant micromotion takes place.

We also find that the area averaged over several Floquet periods (Fig.~\ref{fig:2}c)  becomes more and more peaked at the fine-tuned point as a function of the number of Floquet periods. In the fine-tuned point of dispersionless dynamics, the  time-evolving state remains localized and therefore keeps the quantized response even averaged over many Floquet periods. This effect makes possible to locate  with higher precision the fine-tuned point which is at the "center" of the anomalous phase and where a more robust edge-mode preparation could be achieved \cite{Quelle2017}. 

The derivations hold generally and also apply to a bipartite quadratic lattice, where the tunneling direction is alternated cyclically along the four nearest-neighbors directions~\cite{Rudner2013}.  We numerically verify that the micromotion area behaves qualitatively similar in this case, but we find it to be more peaked around the fine-tuned point as compared to the honeycomb lattice (Fig.~\ref{fig:2}d).
 
\begin{figure}[h]
\includegraphics[width=0.9\linewidth]{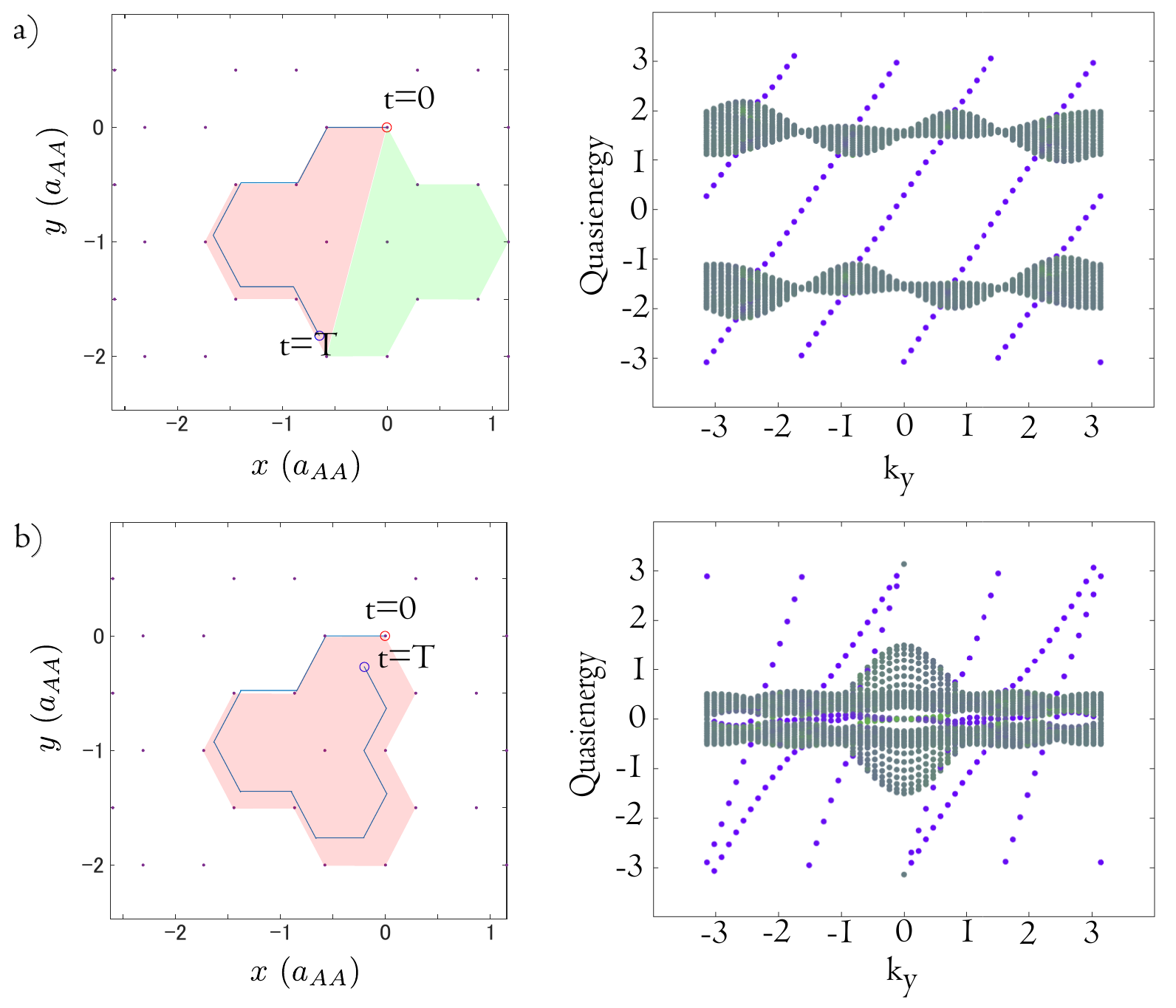}
\caption{{Protocols for higher winding numbers.} a) Dynamics in the bulk for a protocol where tunneling couplings $J_i$ are alternated according to 1313212. The unit of length is the lattice vector $a_{AA}$. The driving frequency $\omega$ is chosen to be perfectly resonant, and a deviation from the fine-tuned condition is given by a $\Delta=0.221J$. {The red shading indicates the area enclosed during one Floquet period at the associated fine-tuned point. The right panel shows the spectrum in a cylinder geometry with gray color for bulk states and purple color for states residing on one of the edges (the states on the other edge of the system are not shown). The spectrum indicates the winding numbers $W_0=W_\pi=4$, which matches with the red area $A=2A_u=\frac{W_{0,\pi}}{2}A_u$. The green shading indicates the area enclosed during the second Floquet period.}  b) Same for the driving sequence 131321213232.  
Here $\omega$ is 0.92 {of} the resonant value, and $\Delta=0$. While the $0$ gap is closed, the spectrum exhibits $W_{\pi}=2\frac{A}{A_u}=6$ {and a corresponding red shaded area $A=3A_u=\frac{W_\pi}{2}A_u$}. } 
\label{fig:3}
\end{figure}

The winding numbers can take arbitrary integer values and values up to $W_g=2$ were previously considered for system with $W_{0}\neq W_\pi$ \cite{Shi2024_W2,PhysRevLett.130.043201Zhang_TuningFloquetTopoBands}. We note that the above relations hold for all anomalous Floquet systems with any winding number $W_{0}=W_\pi>0$ and that these systems can therefore be identified via the micromotion area. 
Furthermore, the relations guide the design of protocols for such new anomalous Floquet phases with $W_{0}=W_\pi\gg 1$: the required larger micromotion areas can be obtained by using protocols with more complex sequences of alternating tunnel couplings along the three directions. We discuss two examples with $W_{0,\pi}=4$ and $W_\pi=6$ in Fig.~\ref{fig:3} and verify the relation $W_{0,\pi}=2\frac{A}{A_u}$ for these protocols close to the fine-tuned point by analysis of the micromotion area and the edge states in the spectra.

Note that a cyclotron orbit was previously detected for a single plaquette with an artificial magnetic field in the effective time-independent Hamiltonian \cite{Aidelsburger2013} and in a synthetic quantum Hall system in mixed dimensions \cite{Chalopin2020,Roell2023}. These measurements do not correspond to the observation of an anomalous phase, because dynamics is induced in those systems by state preparation on one side of an isolated plaquette \cite{Aidelsburger2013} or by an initial momentum transfer along one direction \cite{Chalopin2020,Roell2023}. In these models, the center of mass of a particle localized at $t=0$ does not exhibit any cyclotron dynamics. Finally, in nonlinear photonic systems the area encircled by the center of mass of a soliton was used to distinguish a chiral from a trivial soliton  \cite{Mukherjee2020} and our work provides the theoretical foundation for this phenomenological connection. Note that anomalous phases with $W_0=W_\pi=1$ can be obtained not only by cyclic tunnel modulation as studied here but also by lattice shaking \cite{Wintersperger2020phd}.
\newline

{\it Conclusions and Outlook - }
We introduced a very direct real-space observable as a proxy to detect anomalous topology in general driven two-band systems with cyclic tunnel modulation. Its measurement only requires the observation of the real-space dynamics of a particle prepared on a single lattice site. The real-space detection scheme of the anomalous phase based on the cyclotron orbit is particularly useful for inhomogeneous systems, systems with disorder \cite{Titum2016,Hesse2025}, or at an interface between system parts characterized by different topology. An interesting open question is the generalization of the approach to systems with more than two bands. 
\newline

{\it Acknowledgments - }
The authors thank Nathan Goldman, Lucila Peralta Gavensky, Gonzalo Usaj, Nur Ünal for stimulating discussion and acknowledge support by the
Deutsche Forschungsgemeinschaft (DFG, German Research Foundation) via the Research Unit FOR 5688 (Project No. 521530974) and via the cluster of excellence AIM, EXC 2056 (Project No. 390715994). K. S. acknowledges funding via 'Hamburg Quantum Computing', financed by EU within EFRE and FHH.


\begin{thebibliography}{32}%
\makeatletter
\providecommand \@ifxundefined [1]{%
 \@ifx{#1\undefined}
}%
\providecommand \@ifnum [1]{%
 \ifnum #1\expandafter \@firstoftwo
 \else \expandafter \@secondoftwo
 \fi
}%
\providecommand \@ifx [1]{%
 \ifx #1\expandafter \@firstoftwo
 \else \expandafter \@secondoftwo
 \fi
}%
\providecommand \natexlab [1]{#1}%
\providecommand \enquote  [1]{``#1''}%
\providecommand \bibnamefont  [1]{#1}%
\providecommand \bibfnamefont [1]{#1}%
\providecommand \citenamefont [1]{#1}%
\providecommand \href@noop [0]{\@secondoftwo}%
\providecommand \href [0]{\begingroup \@sanitize@url \@href}%
\providecommand \@href[1]{\@@startlink{#1}\@@href}%
\providecommand \@@href[1]{\endgroup#1\@@endlink}%
\providecommand \@sanitize@url [0]{\catcode `\\12\catcode `\$12\catcode
  `\&12\catcode `\#12\catcode `\^12\catcode `\_12\catcode `\%12\relax}%
\providecommand \@@startlink[1]{}%
\providecommand \@@endlink[0]{}%
\providecommand \url  [0]{\begingroup\@sanitize@url \@url }%
\providecommand \@url [1]{\endgroup\@href {#1}{\urlprefix }}%
\providecommand \urlprefix  [0]{URL }%
\providecommand \Eprint [0]{\href }%
\providecommand \doibase [0]{https://doi.org/}%
\providecommand \selectlanguage [0]{\@gobble}%
\providecommand \bibinfo  [0]{\@secondoftwo}%
\providecommand \bibfield  [0]{\@secondoftwo}%
\providecommand \translation [1]{[#1]}%
\providecommand \BibitemOpen [0]{}%
\providecommand \bibitemStop [0]{}%
\providecommand \bibitemNoStop [0]{.\EOS\space}%
\providecommand \EOS [0]{\spacefactor3000\relax}%
\providecommand \BibitemShut  [1]{\csname bibitem#1\endcsname}%
\let\auto@bib@innerbib\@empty
\bibitem [{\citenamefont {Hasan}\ and\ \citenamefont {Kane}(2010)}]{Hasan2010}%
  \BibitemOpen
  \bibfield  {author} {\bibinfo {author} {\bibfnamefont {M.~Z.}\ \bibnamefont
  {Hasan}}\ and\ \bibinfo {author} {\bibfnamefont {C.~L.}\ \bibnamefont
  {Kane}},\ }\href@noop {} {\bibfield  {journal} {\bibinfo  {journal} {Reviews
  of Modern Physics}\ }\textbf {\bibinfo {volume} {82}},\ \bibinfo {pages}
  {3045} (\bibinfo {year} {2010})}\BibitemShut {NoStop}%
\bibitem [{\citenamefont {Schnyder}\ \emph {et~al.}(2008)\citenamefont
  {Schnyder}, \citenamefont {Ryu}, \citenamefont {Furusaki},\ and\
  \citenamefont {Ludwig}}]{Schnyder2008_classification}%
  \BibitemOpen
  \bibfield  {author} {\bibinfo {author} {\bibfnamefont {A.~P.}\ \bibnamefont
  {Schnyder}}, \bibinfo {author} {\bibfnamefont {S.}~\bibnamefont {Ryu}},
  \bibinfo {author} {\bibfnamefont {A.}~\bibnamefont {Furusaki}},\ and\
  \bibinfo {author} {\bibfnamefont {A.~W.~W.}\ \bibnamefont {Ludwig}},\ }\href
  {https://doi.org/10.1103/PhysRevB.78.195125} {\bibfield  {journal} {\bibinfo
  {journal} {Phys. Rev. B}\ }\textbf {\bibinfo {volume} {78}},\ \bibinfo
  {pages} {195125} (\bibinfo {year} {2008})}\BibitemShut {NoStop}%
\bibitem [{\citenamefont {Kitaev}(2009)}]{kitaev2009}%
  \BibitemOpen
  \bibfield  {author} {\bibinfo {author} {\bibfnamefont {A.}~\bibnamefont
  {Kitaev}},\ }\href {https://doi.org/10.1063/1.3149495} {\bibfield  {journal}
  {\bibinfo  {journal} {AIP Conference Proceedings}\ }\textbf {\bibinfo
  {volume} {1134}},\ \bibinfo {pages} {22} (\bibinfo {year} {2009})},\ \Eprint
  {https://arxiv.org/abs/https://pubs.aip.org/aip/acp/article-pdf/1134/1/22/11584243/22\_1\_online.pdf}
  {https://pubs.aip.org/aip/acp/article-pdf/1134/1/22/11584243/22\_1\_online.pdf}
  \BibitemShut {NoStop}%
\bibitem [{\citenamefont {Ryu}\ \emph {et~al.}(2010)\citenamefont {Ryu},
  \citenamefont {Schnyder}, \citenamefont {Furusaki},\ and\ \citenamefont
  {Ludwig}}]{Ryu_2010}%
  \BibitemOpen
  \bibfield  {author} {\bibinfo {author} {\bibfnamefont {S.}~\bibnamefont
  {Ryu}}, \bibinfo {author} {\bibfnamefont {A.~P.}\ \bibnamefont {Schnyder}},
  \bibinfo {author} {\bibfnamefont {A.}~\bibnamefont {Furusaki}},\ and\
  \bibinfo {author} {\bibfnamefont {A.~W.~W.}\ \bibnamefont {Ludwig}},\ }\href
  {https://doi.org/10.1088/1367-2630/12/6/065010} {\bibfield  {journal}
  {\bibinfo  {journal} {New Journal of Physics}\ }\textbf {\bibinfo {volume}
  {12}},\ \bibinfo {pages} {065010} (\bibinfo {year} {2010})}\BibitemShut
  {NoStop}%
\bibitem [{\citenamefont {Eckardt}(2017)}]{eckardtColloquiumFloquet}%
  \BibitemOpen
  \bibfield  {author} {\bibinfo {author} {\bibfnamefont {A.}~\bibnamefont
  {Eckardt}},\ }\href {https://doi.org/10.1103/RevModPhys.89.011004} {\bibfield
   {journal} {\bibinfo  {journal} {Rev. Mod. Phys.}\ }\textbf {\bibinfo
  {volume} {89}},\ \bibinfo {pages} {011004} (\bibinfo {year}
  {2017})}\BibitemShut {NoStop}%
\bibitem [{\citenamefont {Weitenberg}\ and\ \citenamefont
  {Simonet}(2021)}]{Weitenberg2021}%
  \BibitemOpen
  \bibfield  {author} {\bibinfo {author} {\bibfnamefont {C.}~\bibnamefont
  {Weitenberg}}\ and\ \bibinfo {author} {\bibfnamefont {J.}~\bibnamefont
  {Simonet}},\ }\bibfield  {journal} {\bibinfo  {journal} {Nature Physics}\
  }\href {https://doi.org/10.1038/s41567-021-01316-x}
  {10.1038/s41567-021-01316-x} (\bibinfo {year} {2021})\BibitemShut {NoStop}%
\bibitem [{\citenamefont {Kitagawa}\ \emph {et~al.}(2010)\citenamefont
  {Kitagawa}, \citenamefont {Berg}, \citenamefont {Rudner},\ and\ \citenamefont
  {Demler}}]{Kitagawa2010}%
  \BibitemOpen
  \bibfield  {author} {\bibinfo {author} {\bibfnamefont {T.}~\bibnamefont
  {Kitagawa}}, \bibinfo {author} {\bibfnamefont {E.}~\bibnamefont {Berg}},
  \bibinfo {author} {\bibfnamefont {M.}~\bibnamefont {Rudner}},\ and\ \bibinfo
  {author} {\bibfnamefont {E.}~\bibnamefont {Demler}},\ }\href
  {https://doi.org/10.1103/PhysRevB.82.235114} {\bibfield  {journal} {\bibinfo
  {journal} {Phys. Rev. B}\ }\textbf {\bibinfo {volume} {82}},\ \bibinfo
  {pages} {235114} (\bibinfo {year} {2010})}\BibitemShut {NoStop}%
\bibitem [{\citenamefont {Rudner}\ \emph {et~al.}(2013)\citenamefont {Rudner},
  \citenamefont {Lindner}, \citenamefont {Berg},\ and\ \citenamefont
  {Levin}}]{Rudner2013}%
  \BibitemOpen
  \bibfield  {author} {\bibinfo {author} {\bibfnamefont {M.~S.}\ \bibnamefont
  {Rudner}}, \bibinfo {author} {\bibfnamefont {N.~H.}\ \bibnamefont {Lindner}},
  \bibinfo {author} {\bibfnamefont {E.}~\bibnamefont {Berg}},\ and\ \bibinfo
  {author} {\bibfnamefont {M.}~\bibnamefont {Levin}},\ }\href
  {https://doi.org/10.1103/PhysRevX.3.031005} {\bibfield  {journal} {\bibinfo
  {journal} {Phys. Rev. X}\ }\textbf {\bibinfo {volume} {3}},\ \bibinfo {pages}
  {031005} (\bibinfo {year} {2013})}\BibitemShut {NoStop}%
\bibitem [{\citenamefont
  {G{\'{o}}mez-Le{\'{o}}n}(2024)}]{GomezLeon2024anomalousfloquet}%
  \BibitemOpen
  \bibfield  {author} {\bibinfo {author} {\bibfnamefont {{\'{A}}.}~\bibnamefont
  {G{\'{o}}mez-Le{\'{o}}n}},\ }\href
  {https://doi.org/10.22331/q-2024-11-13-1522} {\bibfield  {journal} {\bibinfo
  {journal} {{Quantum}}\ }\textbf {\bibinfo {volume} {8}},\ \bibinfo {pages}
  {1522} (\bibinfo {year} {2024})}\BibitemShut {NoStop}%
\bibitem [{\citenamefont {Maczewsky}\ \emph {et~al.}(2017)\citenamefont
  {Maczewsky}, \citenamefont {Zeuner}, \citenamefont {Nolte},\ and\
  \citenamefont {Szameit}}]{Maczewsky2017}%
  \BibitemOpen
  \bibfield  {author} {\bibinfo {author} {\bibfnamefont {L.~J.}\ \bibnamefont
  {Maczewsky}}, \bibinfo {author} {\bibfnamefont {J.~M.}\ \bibnamefont
  {Zeuner}}, \bibinfo {author} {\bibfnamefont {S.}~\bibnamefont {Nolte}},\ and\
  \bibinfo {author} {\bibfnamefont {A.}~\bibnamefont {Szameit}},\ }\href
  {https://doi.org/10.1038/ncomms13756} {\bibfield  {journal} {\bibinfo
  {journal} {Nature Communications}\ }\textbf {\bibinfo {volume} {8}},\
  \bibinfo {pages} {13756} (\bibinfo {year} {2017})}\BibitemShut {NoStop}%
\bibitem [{\citenamefont {Mukherjee}\ \emph {et~al.}(2017)\citenamefont
  {Mukherjee}, \citenamefont {Spracklen}, \citenamefont {Valiente},
  \citenamefont {Andersson}, \citenamefont {{\"{O}}hberg}, \citenamefont
  {Goldman},\ and\ \citenamefont {Thomson}}]{Mukherjee2017}%
  \BibitemOpen
  \bibfield  {author} {\bibinfo {author} {\bibfnamefont {S.}~\bibnamefont
  {Mukherjee}}, \bibinfo {author} {\bibfnamefont {A.}~\bibnamefont
  {Spracklen}}, \bibinfo {author} {\bibfnamefont {M.}~\bibnamefont {Valiente}},
  \bibinfo {author} {\bibfnamefont {E.}~\bibnamefont {Andersson}}, \bibinfo
  {author} {\bibfnamefont {P.}~\bibnamefont {{\"{O}}hberg}}, \bibinfo {author}
  {\bibfnamefont {N.}~\bibnamefont {Goldman}},\ and\ \bibinfo {author}
  {\bibfnamefont {R.~R.}\ \bibnamefont {Thomson}},\ }\href
  {https://doi.org/10.1038/ncomms13918} {\bibfield  {journal} {\bibinfo
  {journal} {Nature Communications}\ }\textbf {\bibinfo {volume} {8}},\
  \bibinfo {pages} {1} (\bibinfo {year} {2017})}\BibitemShut {NoStop}%
\bibitem [{\citenamefont {Peng}\ \emph {et~al.}(2016)\citenamefont {Peng},
  \citenamefont {Qin}, \citenamefont {Zhao}, \citenamefont {Shen},
  \citenamefont {Xu}, \citenamefont {Bao}, \citenamefont {Jia},\ and\
  \citenamefont {Zhu}}]{Peng2016}%
  \BibitemOpen
  \bibfield  {author} {\bibinfo {author} {\bibfnamefont {Y.-G.}\ \bibnamefont
  {Peng}}, \bibinfo {author} {\bibfnamefont {C.-Z.}\ \bibnamefont {Qin}},
  \bibinfo {author} {\bibfnamefont {D.-G.}\ \bibnamefont {Zhao}}, \bibinfo
  {author} {\bibfnamefont {Y.-X.}\ \bibnamefont {Shen}}, \bibinfo {author}
  {\bibfnamefont {X.-Y.}\ \bibnamefont {Xu}}, \bibinfo {author} {\bibfnamefont
  {M.}~\bibnamefont {Bao}}, \bibinfo {author} {\bibfnamefont {H.}~\bibnamefont
  {Jia}},\ and\ \bibinfo {author} {\bibfnamefont {X.-F.}\ \bibnamefont {Zhu}},\
  }\href {https://doi.org/10.1038/ncomms13368} {\bibfield  {journal} {\bibinfo
  {journal} {Nature Communications}\ }\textbf {\bibinfo {volume} {7}},\
  \bibinfo {pages} {13368} (\bibinfo {year} {2016})}\BibitemShut {NoStop}%
\bibitem [{\citenamefont {Wintersperger}\ \emph {et~al.}(2020)\citenamefont
  {Wintersperger}, \citenamefont {Braun}, \citenamefont {{\"U}nal},
  \citenamefont {Eckardt}, \citenamefont {Liberto}, \citenamefont {Goldman},
  \citenamefont {Bloch},\ and\ \citenamefont
  {Aidelsburger}}]{Wintersperger2020}%
  \BibitemOpen
  \bibfield  {author} {\bibinfo {author} {\bibfnamefont {K.}~\bibnamefont
  {Wintersperger}}, \bibinfo {author} {\bibfnamefont {C.}~\bibnamefont
  {Braun}}, \bibinfo {author} {\bibfnamefont {F.~N.}\ \bibnamefont {{\"U}nal}},
  \bibinfo {author} {\bibfnamefont {A.}~\bibnamefont {Eckardt}}, \bibinfo
  {author} {\bibfnamefont {M.~D.}\ \bibnamefont {Liberto}}, \bibinfo {author}
  {\bibfnamefont {N.}~\bibnamefont {Goldman}}, \bibinfo {author} {\bibfnamefont
  {I.}~\bibnamefont {Bloch}},\ and\ \bibinfo {author} {\bibfnamefont
  {M.}~\bibnamefont {Aidelsburger}},\ }\href
  {https://doi.org/10.1038/s41567-020-0949-y} {\bibfield  {journal} {\bibinfo
  {journal} {Nature Physics}\ }\textbf {\bibinfo {volume} {16}},\ \bibinfo
  {pages} {1058} (\bibinfo {year} {2020})}\BibitemShut {NoStop}%
\bibitem [{\citenamefont {Braun}\ \emph {et~al.}(2024)\citenamefont {Braun},
  \citenamefont {Saint-Jalm}, \citenamefont {Hesse}, \citenamefont {Arceri},
  \citenamefont {Bloch},\ and\ \citenamefont {Aidelsburger}}]{Braun2024}%
  \BibitemOpen
  \bibfield  {author} {\bibinfo {author} {\bibfnamefont {C.}~\bibnamefont
  {Braun}}, \bibinfo {author} {\bibfnamefont {R.}~\bibnamefont {Saint-Jalm}},
  \bibinfo {author} {\bibfnamefont {A.}~\bibnamefont {Hesse}}, \bibinfo
  {author} {\bibfnamefont {J.}~\bibnamefont {Arceri}}, \bibinfo {author}
  {\bibfnamefont {I.}~\bibnamefont {Bloch}},\ and\ \bibinfo {author}
  {\bibfnamefont {M.}~\bibnamefont {Aidelsburger}},\ }\href
  {https://doi.org/10.1038/s41567-024-02506-z} {\bibfield  {journal} {\bibinfo
  {journal} {Nature Physics}\ } (\bibinfo {year} {2024})}\BibitemShut {NoStop}%
\bibitem [{\citenamefont {\"Unal}\ \emph {et~al.}(2019)\citenamefont {\"Unal},
  \citenamefont {Seradjeh},\ and\ \citenamefont {Eckardt}}]{Uenal2019}%
  \BibitemOpen
  \bibfield  {author} {\bibinfo {author} {\bibfnamefont {F.~N.}\ \bibnamefont
  {\"Unal}}, \bibinfo {author} {\bibfnamefont {B.}~\bibnamefont {Seradjeh}},\
  and\ \bibinfo {author} {\bibfnamefont {A.}~\bibnamefont {Eckardt}},\ }\href
  {https://doi.org/10.1103/PhysRevLett.122.253601} {\bibfield  {journal}
  {\bibinfo  {journal} {Phys. Rev. Lett.}\ }\textbf {\bibinfo {volume} {122}},\
  \bibinfo {pages} {253601} (\bibinfo {year} {2019})}\BibitemShut {NoStop}%
\bibitem [{\citenamefont {Aidelsburger}\ \emph {et~al.}(2015)\citenamefont
  {Aidelsburger}, \citenamefont {Lohse}, \citenamefont {Schweizer},
  \citenamefont {Atala}, \citenamefont {Barreiro}, \citenamefont
  {Nascimb{\`{e}}ne}, \citenamefont {Cooper}, \citenamefont {Bloch},\ and\
  \citenamefont {Goldman}}]{Aidelsburger2015}%
  \BibitemOpen
  \bibfield  {author} {\bibinfo {author} {\bibfnamefont {M.}~\bibnamefont
  {Aidelsburger}}, \bibinfo {author} {\bibfnamefont {M.}~\bibnamefont {Lohse}},
  \bibinfo {author} {\bibfnamefont {C.}~\bibnamefont {Schweizer}}, \bibinfo
  {author} {\bibfnamefont {M.}~\bibnamefont {Atala}}, \bibinfo {author}
  {\bibfnamefont {J.~T.}\ \bibnamefont {Barreiro}}, \bibinfo {author}
  {\bibfnamefont {S.}~\bibnamefont {Nascimb{\`{e}}ne}}, \bibinfo {author}
  {\bibfnamefont {N.~R.}\ \bibnamefont {Cooper}}, \bibinfo {author}
  {\bibfnamefont {I.}~\bibnamefont {Bloch}},\ and\ \bibinfo {author}
  {\bibfnamefont {N.}~\bibnamefont {Goldman}},\ }\href
  {https://doi.org/10.1038/NPHYS3171} {\bibfield  {journal} {\bibinfo
  {journal} {Nature Phys.}\ }\textbf {\bibinfo {volume} {11}},\ \bibinfo
  {pages} {162} (\bibinfo {year} {2015})}\BibitemShut {NoStop}%
\bibitem [{\citenamefont {Chalopin}\ \emph {et~al.}(2020)\citenamefont
  {Chalopin}, \citenamefont {Satoor}, \citenamefont {Evrard}, \citenamefont
  {Makhalov}, \citenamefont {Dalibard}, \citenamefont {Lopes},\ and\
  \citenamefont {Nascimbene}}]{Chalopin2020}%
  \BibitemOpen
  \bibfield  {author} {\bibinfo {author} {\bibfnamefont {T.}~\bibnamefont
  {Chalopin}}, \bibinfo {author} {\bibfnamefont {T.}~\bibnamefont {Satoor}},
  \bibinfo {author} {\bibfnamefont {A.}~\bibnamefont {Evrard}}, \bibinfo
  {author} {\bibfnamefont {V.}~\bibnamefont {Makhalov}}, \bibinfo {author}
  {\bibfnamefont {J.}~\bibnamefont {Dalibard}}, \bibinfo {author}
  {\bibfnamefont {R.}~\bibnamefont {Lopes}},\ and\ \bibinfo {author}
  {\bibfnamefont {S.}~\bibnamefont {Nascimbene}},\ }\href
  {https://doi.org/10.1038/s41567-020-0942-5} {\bibfield  {journal} {\bibinfo
  {journal} {Nature Physics}\ }\textbf {\bibinfo {volume} {16}},\ \bibinfo
  {pages} {1017} (\bibinfo {year} {2020})}\BibitemShut {NoStop}%
\bibitem [{\citenamefont {Bianco}\ and\ \citenamefont
  {Resta}(2011)}]{BiancoResta2011}%
  \BibitemOpen
  \bibfield  {author} {\bibinfo {author} {\bibfnamefont {R.}~\bibnamefont
  {Bianco}}\ and\ \bibinfo {author} {\bibfnamefont {R.}~\bibnamefont {Resta}},\
  }\href {https://doi.org/10.1103/PhysRevB.84.241106} {\bibfield  {journal}
  {\bibinfo  {journal} {Phys. Rev. B}\ }\textbf {\bibinfo {volume} {84}},\
  \bibinfo {pages} {241106} (\bibinfo {year} {2011})}\BibitemShut {NoStop}%
\bibitem [{\citenamefont {Peralta~Gavensky}\ \emph {et~al.}(2025)\citenamefont
  {Peralta~Gavensky}, \citenamefont {Usaj},\ and\ \citenamefont
  {Goldman}}]{gavensky2024}%
  \BibitemOpen
  \bibfield  {author} {\bibinfo {author} {\bibfnamefont {L.}~\bibnamefont
  {Peralta~Gavensky}}, \bibinfo {author} {\bibfnamefont {G.}~\bibnamefont
  {Usaj}},\ and\ \bibinfo {author} {\bibfnamefont {N.}~\bibnamefont
  {Goldman}},\ }\href {https://doi.org/10.1103/b3pw-my97} {\bibfield  {journal}
  {\bibinfo  {journal} {Phys. Rev. X}\ }\textbf {\bibinfo {volume} {15}},\
  \bibinfo {pages} {031067} (\bibinfo {year} {2025})}\BibitemShut {NoStop}%
\bibitem [{\citenamefont {Cardano}\ \emph {et~al.}(2017)\citenamefont
  {Cardano}, \citenamefont {D’Errico}, \citenamefont {Dauphin}, \citenamefont
  {Maffei}, \citenamefont {Piccirillo}, \citenamefont {de~Lisio}, \citenamefont
  {De~Filippis}, \citenamefont {Cataudella}, \citenamefont {Santamato},
  \citenamefont {Marrucci}, \citenamefont {Lewenstein},\ and\ \citenamefont
  {Massignan}}]{Cardano2017}%
  \BibitemOpen
  \bibfield  {author} {\bibinfo {author} {\bibfnamefont {F.}~\bibnamefont
  {Cardano}}, \bibinfo {author} {\bibfnamefont {A.}~\bibnamefont {D’Errico}},
  \bibinfo {author} {\bibfnamefont {A.}~\bibnamefont {Dauphin}}, \bibinfo
  {author} {\bibfnamefont {M.}~\bibnamefont {Maffei}}, \bibinfo {author}
  {\bibfnamefont {B.}~\bibnamefont {Piccirillo}}, \bibinfo {author}
  {\bibfnamefont {C.}~\bibnamefont {de~Lisio}}, \bibinfo {author}
  {\bibfnamefont {G.}~\bibnamefont {De~Filippis}}, \bibinfo {author}
  {\bibfnamefont {V.}~\bibnamefont {Cataudella}}, \bibinfo {author}
  {\bibfnamefont {E.}~\bibnamefont {Santamato}}, \bibinfo {author}
  {\bibfnamefont {L.}~\bibnamefont {Marrucci}}, \bibinfo {author}
  {\bibfnamefont {M.}~\bibnamefont {Lewenstein}},\ and\ \bibinfo {author}
  {\bibfnamefont {P.}~\bibnamefont {Massignan}},\ }\href
  {https://doi.org/10.1038/ncomms15516} {\bibfield  {journal} {\bibinfo
  {journal} {Nature Communications}\ }\textbf {\bibinfo {volume} {8}},\
  \bibinfo {pages} {15516} (\bibinfo {year} {2017})}\BibitemShut {NoStop}%
\bibitem [{\citenamefont {Maffei}\ \emph {et~al.}(2018)\citenamefont {Maffei},
  \citenamefont {Dauphin}, \citenamefont {Cardano}, \citenamefont
  {Lewenstein},\ and\ \citenamefont {Massignan}}]{Maffei2018}%
  \BibitemOpen
  \bibfield  {author} {\bibinfo {author} {\bibfnamefont {M.}~\bibnamefont
  {Maffei}}, \bibinfo {author} {\bibfnamefont {A.}~\bibnamefont {Dauphin}},
  \bibinfo {author} {\bibfnamefont {F.}~\bibnamefont {Cardano}}, \bibinfo
  {author} {\bibfnamefont {M.}~\bibnamefont {Lewenstein}},\ and\ \bibinfo
  {author} {\bibfnamefont {P.}~\bibnamefont {Massignan}},\ }\href
  {https://doi.org/10.1088/1367-2630/aa9d4c} {\bibfield  {journal} {\bibinfo
  {journal} {New Journal of Physics}\ }\textbf {\bibinfo {volume} {20}},\
  \bibinfo {pages} {013023} (\bibinfo {year} {2018})}\BibitemShut {NoStop}%
\bibitem [{\citenamefont {Vanderbilt}(2018)}]{Vanderbilt_2018}%
  \BibitemOpen
  \bibfield  {author} {\bibinfo {author} {\bibfnamefont {D.}~\bibnamefont
  {Vanderbilt}},\ }\bibinfo {title} {Berry phases and curvatures},\ in\
  \href@noop {} {\emph {\bibinfo {booktitle} {Berry Phases in Electronic
  Structure Theory: Electric Polarization, Orbital Magnetization and
  Topological Insulators}}}\ (\bibinfo  {publisher} {Cambridge University
  Press},\ \bibinfo {year} {2018})\ p.\ \bibinfo {pages} {117}\BibitemShut
  {NoStop}%
\bibitem [{\citenamefont {Mart\'{\i}nez}\ and\ \citenamefont
  {\"Unal}(2023)}]{Martinez2023}%
  \BibitemOpen
  \bibfield  {author} {\bibinfo {author} {\bibfnamefont {M.~F.}\ \bibnamefont
  {Mart\'{\i}nez}}\ and\ \bibinfo {author} {\bibfnamefont {F.~N.}\ \bibnamefont
  {\"Unal}},\ }\href {https://doi.org/10.1103/PhysRevA.108.063314} {\bibfield
  {journal} {\bibinfo  {journal} {Phys. Rev. A}\ }\textbf {\bibinfo {volume}
  {108}},\ \bibinfo {pages} {063314} (\bibinfo {year} {2023})}\BibitemShut
  {NoStop}%
\bibitem [{\citenamefont {Quelle}\ \emph {et~al.}(2017)\citenamefont {Quelle},
  \citenamefont {Weitenberg}, \citenamefont {Sengstock},\ and\ \citenamefont
  {Smith}}]{Quelle2017}%
  \BibitemOpen
  \bibfield  {author} {\bibinfo {author} {\bibfnamefont {A.}~\bibnamefont
  {Quelle}}, \bibinfo {author} {\bibfnamefont {C.}~\bibnamefont {Weitenberg}},
  \bibinfo {author} {\bibfnamefont {K.}~\bibnamefont {Sengstock}},\ and\
  \bibinfo {author} {\bibfnamefont {C.~M.}\ \bibnamefont {Smith}},\ }\href
  {https://doi.org/10.1088/1367-2630/aa8646} {\bibfield  {journal} {\bibinfo
  {journal} {New Journal of Physics}\ }\textbf {\bibinfo {volume} {19}},\
  \bibinfo {pages} {113010} (\bibinfo {year} {2017})}\BibitemShut {NoStop}%
\bibitem [{\citenamefont {Shi}\ \emph {et~al.}(2024)\citenamefont {Shi},
  \citenamefont {Zhang},\ and\ \citenamefont {Zhang}}]{Shi2024_W2}%
  \BibitemOpen
  \bibfield  {author} {\bibinfo {author} {\bibfnamefont {K.}~\bibnamefont
  {Shi}}, \bibinfo {author} {\bibfnamefont {X.}~\bibnamefont {Zhang}},\ and\
  \bibinfo {author} {\bibfnamefont {W.}~\bibnamefont {Zhang}},\ }\href
  {https://doi.org/10.1103/PhysRevA.109.013324} {\bibfield  {journal} {\bibinfo
   {journal} {Phys. Rev. A}\ }\textbf {\bibinfo {volume} {109}},\ \bibinfo
  {pages} {013324} (\bibinfo {year} {2024})}\BibitemShut {NoStop}%
\bibitem [{\citenamefont {Zhang}\ \emph {et~al.}(2023)\citenamefont {Zhang},
  \citenamefont {Yi}, \citenamefont {Zhang}, \citenamefont {Jiao},
  \citenamefont {Shi}, \citenamefont {Yuan}, \citenamefont {Zhang},
  \citenamefont {Liu}, \citenamefont {Chen},\ and\ \citenamefont
  {Pan}}]{PhysRevLett.130.043201Zhang_TuningFloquetTopoBands}%
  \BibitemOpen
  \bibfield  {author} {\bibinfo {author} {\bibfnamefont {J.-Y.}\ \bibnamefont
  {Zhang}}, \bibinfo {author} {\bibfnamefont {C.-R.}\ \bibnamefont {Yi}},
  \bibinfo {author} {\bibfnamefont {L.}~\bibnamefont {Zhang}}, \bibinfo
  {author} {\bibfnamefont {R.-H.}\ \bibnamefont {Jiao}}, \bibinfo {author}
  {\bibfnamefont {K.-Y.}\ \bibnamefont {Shi}}, \bibinfo {author} {\bibfnamefont
  {H.}~\bibnamefont {Yuan}}, \bibinfo {author} {\bibfnamefont {W.}~\bibnamefont
  {Zhang}}, \bibinfo {author} {\bibfnamefont {X.-J.}\ \bibnamefont {Liu}},
  \bibinfo {author} {\bibfnamefont {S.}~\bibnamefont {Chen}},\ and\ \bibinfo
  {author} {\bibfnamefont {J.-W.}\ \bibnamefont {Pan}},\ }\href
  {https://doi.org/10.1103/PhysRevLett.130.043201} {\bibfield  {journal}
  {\bibinfo  {journal} {Phys. Rev. Lett.}\ }\textbf {\bibinfo {volume} {130}},\
  \bibinfo {pages} {043201} (\bibinfo {year} {2023})}\BibitemShut {NoStop}%
\bibitem [{\citenamefont {Aidelsburger}\ \emph {et~al.}(2013)\citenamefont
  {Aidelsburger}, \citenamefont {Atala}, \citenamefont {Lohse}, \citenamefont
  {Barreiro}, \citenamefont {Paredes},\ and\ \citenamefont
  {Bloch}}]{Aidelsburger2013}%
  \BibitemOpen
  \bibfield  {author} {\bibinfo {author} {\bibfnamefont {M.}~\bibnamefont
  {Aidelsburger}}, \bibinfo {author} {\bibfnamefont {M.}~\bibnamefont {Atala}},
  \bibinfo {author} {\bibfnamefont {M.}~\bibnamefont {Lohse}}, \bibinfo
  {author} {\bibfnamefont {J.~T.}\ \bibnamefont {Barreiro}}, \bibinfo {author}
  {\bibfnamefont {B.}~\bibnamefont {Paredes}},\ and\ \bibinfo {author}
  {\bibfnamefont {I.}~\bibnamefont {Bloch}},\ }\href
  {https://doi.org/10.1103/PhysRevLett.111.185301} {\bibfield  {journal}
  {\bibinfo  {journal} {Phys. Rev. Lett.}\ }\textbf {\bibinfo {volume} {111}},\
  \bibinfo {pages} {185301} (\bibinfo {year} {2013})}\BibitemShut {NoStop}%
\bibitem [{\citenamefont {Roell}\ \emph {et~al.}(2023)\citenamefont {Roell},
  \citenamefont {Laskar}, \citenamefont {Huybrechts},\ and\ \citenamefont
  {Weitz}}]{Roell2023}%
  \BibitemOpen
  \bibfield  {author} {\bibinfo {author} {\bibfnamefont {R.~V.}\ \bibnamefont
  {Roell}}, \bibinfo {author} {\bibfnamefont {A.~W.}\ \bibnamefont {Laskar}},
  \bibinfo {author} {\bibfnamefont {F.~R.}\ \bibnamefont {Huybrechts}},\ and\
  \bibinfo {author} {\bibfnamefont {M.}~\bibnamefont {Weitz}},\ }\href
  {https://doi.org/10.1103/PhysRevA.107.043302} {\bibfield  {journal} {\bibinfo
   {journal} {Phys. Rev. A}\ }\textbf {\bibinfo {volume} {107}},\ \bibinfo
  {pages} {043302} (\bibinfo {year} {2023})}\BibitemShut {NoStop}%
\bibitem [{\citenamefont {Mukherjee}\ and\ \citenamefont
  {Rechtsman}(2020)}]{Mukherjee2020}%
  \BibitemOpen
  \bibfield  {author} {\bibinfo {author} {\bibfnamefont {S.}~\bibnamefont
  {Mukherjee}}\ and\ \bibinfo {author} {\bibfnamefont {M.~C.}\ \bibnamefont
  {Rechtsman}},\ }\href {https://doi.org/10.1126/science.aba8725} {\bibfield
  {journal} {\bibinfo  {journal} {Science}\ }\textbf {\bibinfo {volume}
  {368}},\ \bibinfo {pages} {856} (\bibinfo {year} {2020})}\BibitemShut
  {NoStop}%
\bibitem [{\citenamefont {Wintersperger}(2020)}]{Wintersperger2020phd}%
  \BibitemOpen
  \bibfield  {author} {\bibinfo {author} {\bibfnamefont {K.}~\bibnamefont
  {Wintersperger}},\ }\emph {\bibinfo {title} {{Realization of Floquet
  topological systems with ultracold atoms in optical honeycomb lattices}}},\
  \href@noop {} {Ph.D. thesis},\ \bibinfo  {school}
  {Ludwig-Maximilians-Universit{\"{a}}t M{\"{u}}nchen} (\bibinfo {year}
  {2020})\BibitemShut {NoStop}%
\bibitem [{\citenamefont {Titum}\ \emph {et~al.}(2016)\citenamefont {Titum},
  \citenamefont {Berg}, \citenamefont {Rudner}, \citenamefont {Refael},\ and\
  \citenamefont {Lindner}}]{Titum2016}%
  \BibitemOpen
  \bibfield  {author} {\bibinfo {author} {\bibfnamefont {P.}~\bibnamefont
  {Titum}}, \bibinfo {author} {\bibfnamefont {E.}~\bibnamefont {Berg}},
  \bibinfo {author} {\bibfnamefont {M.~S.}\ \bibnamefont {Rudner}}, \bibinfo
  {author} {\bibfnamefont {G.}~\bibnamefont {Refael}},\ and\ \bibinfo {author}
  {\bibfnamefont {N.~H.}\ \bibnamefont {Lindner}},\ }\href
  {https://doi.org/10.1103/PhysRevX.6.021013} {\bibfield  {journal} {\bibinfo
  {journal} {Phys. Rev. X}\ }\textbf {\bibinfo {volume} {6}},\ \bibinfo {pages}
  {021013} (\bibinfo {year} {2016})}\BibitemShut {NoStop}%
\bibitem [{\citenamefont {Hesse}\ \emph {et~al.}(2025)\citenamefont {Hesse},
  \citenamefont {Arceri}, \citenamefont {Hornung}, \citenamefont {Braun},\ and\
  \citenamefont {Aidelsburger}}]{Hesse2025}%
  \BibitemOpen
  \bibfield  {author} {\bibinfo {author} {\bibfnamefont {A.}~\bibnamefont
  {Hesse}}, \bibinfo {author} {\bibfnamefont {J.}~\bibnamefont {Arceri}},
  \bibinfo {author} {\bibfnamefont {M.}~\bibnamefont {Hornung}}, \bibinfo
  {author} {\bibfnamefont {C.}~\bibnamefont {Braun}},\ and\ \bibinfo {author}
  {\bibfnamefont {M.}~\bibnamefont {Aidelsburger}},\ }\href
  {https://arxiv.org/abs/2508.20154} {\bibinfo {title} {Probing disorder-driven
  topological phase transitions via topological edge modes with ultracold atoms
  in floquet-engineered honeycomb lattices}} (\bibinfo {year} {2025}),\ \Eprint
  {https://arxiv.org/abs/2508.20154} {arXiv:2508.20154} \BibitemShut {NoStop}%
\end{thebibliography}
%

\newpage
\section*{End-Matter}
\subsection{Full calculation}
\label{Full_Calculation}
Because the Berry curvature is opposite for different sublattices, both the velocity and the displacement have an opposite sign, resulting in $A_0=A_1$. The distinction between the areas $A_{0,1}$ obtained by starting from different sublattices $s=0,1$ is purely for mathematical convenience. Starting from Eq.~(\ref{Area_with_operators}) and using the definitions of $B$ and $C$ after Eq.~(\ref{main_equation}), we write
\begin{equation}
    2A=A_0+A_1=i\sum_{s=0}^1\int_0^T \mathrm{d}t~ \mathrm{Im}\bigl[\bra{s}\overline{B}\ket{s}\bigr]\times \bra{s}\overline{C}\ket{s},
\end{equation}
where $\overline{B},~\overline{C}$ indicates average over the BZ.

We replace the product of the mean values with the mean value of the product minus the correlations:
\begin{equation}
\begin{aligned}
    2A=i\sum_{s=0}^1\int_0^T\mathrm{d}t~&\overline{\mathrm{Im}\bigl[\bra{s}B\ket{s}\bigr]\times\bra{s}C\ket{s}}-\\ &\overline{\mathrm{Im}\bigl[\bra{s}\delta B\ket{s}\bigr]\times\bra{s}\delta C\ket{s}}
    \label{deviations}
    \end{aligned}
\end{equation}
where $B=\overline{B}+\delta B,~C=\overline{C}+\delta C$.

Because $0=\nabla~1=\nabla ({U}_g^\dagger {U}_g)=\nabla {U}_g^\dagger {U}_g+{U}_g^\dagger \nabla {U}_g$ it follows that ${U}_g^\dagger\nabla {U}_g=-\nabla {U}_g^\dagger {U}_g = -({U}_g^\dagger\nabla {U}_g)^\dagger$, i.e. ${U}_g^\dagger\nabla {U}_g$ is anti-Hermitian, and therefore $\bra{s}C\ket{s}$ is a purely complex number. Therefore:
\begin{equation}
\begin{aligned}
2A=&~i\sum_{s=0}^1\int_0^T\mathrm{d}t~\overline{\mathrm{Im}\bigl[\bra{s}B\ket{s}\bigr]\times\bra{s}C\ket{s}}\\
&-\overline{\mathrm{Im}\bigl[\bra{j}\delta B\ket{j}\bigr]\times\bra{s}\delta C\ket{s}}\\
=&\sum_{s=0}^1\int_0^T\mathrm{d}t~\overline{\mathrm{Re}\bigl[\bra{s}B\ket{s}\times\bra{s}C\ket{s}}\bigr]+T_\mathrm{corr}
\end{aligned}
\label{real_value_expression}
\end{equation}
where $T_\mathrm{corr}=-i\sum_{s=0}^{1}\int_0^T\overline{\mathrm{Im}\bigl[\bra{s}\delta B\ket{s}\bigr]\times\bra{s}\delta C\ket{s}}$ is the position-velocity correlation term.

We compare now the following terms:
\begin{equation}
\begin{aligned}
    a_d=&~\mathrm{Re}\bigl[{\bra{s}B\ket{s}\times\bra{s}C\ket{s}}\bigr]\\
    a_m=&~\mathrm{Re}\bigl[{\bra{s}B\ket{1-s}\times\bra{1-s}C\ket{s}\bigr]}
\end{aligned}
\end{equation}

We demonstrate that $a_d=a_m$. We consider the matrix $M=\partial_t {U}_g^\dagger {U}_g$ such that $\partial_t {U}_g^\dagger=M{U}_g^\dagger$. Then it also holds $B=MC$.

We decompose $M$ as: $M=
  \left( {\begin{array}{cc}
    m_{00} & m_{01} \\
    m_{10} & m_{11} \\
  \end{array} } \right)$

The terms proportional to $m_{11}$ and $m_{00}$ do not contribute neither to $a_d$ nor to $a_m$.

We consider the term $\bra{0}B^x\ket{0}\bra{0}C^y\ket{0}-\bra{0}B^y\ket{0}\bra{0}C^x\ket{0}$ in $a_d$. The part proportional to $m_{00}$ becomes:
\begin{equation}\begin{aligned}
&\bra{0}C^x\ket{0}\bra{0}C^y\ket{0}-\bra{0}C^y\ket{0}\bra{0}C^y\ket{0}=0\end{aligned}
\end{equation}

The part of $\mathrm{Re}\bigl[\bra{0}B^x\ket{1}\bra{1}C^y\ket{0}-\bra{0}B^y\ket{1}\bra{1}C^y\ket{0}\bigr]$ in $a_m$ proportional to $m_{00}$ is also zero:
\begin{equation}\begin{aligned}
&\mathrm{Re}\Bigl[\bra{0}C^x\ket{1}\bra{1}C^y\ket{0}-\bra{0}C^y\ket{1}\bra{1}C^x\ket{0}\Bigr]=\\
&\mathrm{Re}\Bigl[\bra{0}C^x\ket{1}\bra{1}C^y\ket{0}-(\bra{1}C^y\ket{0})^*(\bra{0}C^y\ket{1})^*\Bigr]
\end{aligned}
\end{equation}
because the real part of this difference is zero. The same holds for $m_{11}$. 

We consider now terms proportional to $m_{01}$, and demonstrate that they are the same in $a_d$ and $a_m$. We note that this is the only part which works only for two-band models.

Considering $\bra{0}M{B}_x\ket{0}\bra{0}{C}_y\ket{0}$ (in $a_d$) and $-\bra{0}M{B}_y\ket{1}\bra{1}{C}_x\ket{0}$ (in $a_m$; notice the minus sign because of the opposite ordering of $x$ and $y$).
We obtain:
\begin{equation}
\begin{aligned}
a_d:~~~    \bra{0}M{B}_x\ket{0}\bra{0}{C}_y\ket{0}&\xrightarrow{}m_{01}\bra{1}{C}_x\ket{0}\bra{0}{C}_y\ket{0}\\
a_m:\; -\bra{0}M{B}_y\ket{1}\bra{1}{C}_x\ket{0}&\xrightarrow{}-m_{01}\bra{1}{C}_y\ket{1}\bra{1}{C}_x\ket{0}
\end{aligned}    
\end{equation}

The two terms are equal because $\bra{0}{C}_y\ket{0}=-\bra{1}{C}_y\ket{1}$ (that is because the sum of the Berry curvature over all bands is zero).

A similar reasoning can be repeated for $m_{10}$, demonstrating that $a_s=a_m$.

One can then write: 
\begin{equation}
    \begin{aligned}
    a_d=\frac{1}{2}(a_d+a_m)=&\frac{1}{2}\mathrm{Re\Bigl[}~\bra{s}B\ket{s'}\times\bra{s'}C\ket{s}\Bigr]=\\
    &\frac{1}{2}\mathrm{Re\Bigl[}~\bra{s}B\times C\ket{s}\Bigr]=\\
    &\frac{1}{2}\mathrm{Re~\Bigl[}\mathrm{Tr}~B\times C~\Bigr]=\\
    \end{aligned}
\end{equation}

Hence we write:
\begin{equation}
    \begin{aligned}
        2A=\frac{1}{2}\int_0^T~\;\mathrm{d}t~\mathrm{Re}\Bigl[\mathrm{Tr}~\overline{B\times C}\Bigr]~~+~T_\mathrm{corr}
    \end{aligned}
\end{equation}

Finally, comparing with Eq.~(\ref{WindingN_definition_2}) we get:
\begin{equation}\begin{aligned}
    2A&=A_uW_g-\frac{A_u}{8\pi^2}\int_T^{2T} \mathrm{d}t \int_{BZ}\mathrm{d}{\mathbf{k}}~\mathrm{Re~(Tr}\Bigl[B\times C\Bigr])\\
    &-\sum_{s=0}^1\int_0^{T}\mathrm{d}t~\overline{\mathrm{Im}\bigl[\bra{s}\delta B\ket{s}\bigr]\times\bra{s}\delta C\ket{s}}
    \end{aligned}
\end{equation}
Which demonstrates Eq.~(\ref{main_equation}).

\subsection{Fine-tuned point}
\label{supp:fine-tuned}
Here we calculate the band-flattening term and the velocity-position correlation term at the fine-tuned point (where the dynamics is completely dispersionless), and show that they are zero, making $W_g=2A$ hold.

Let's assume that after a certain number of steps at time $t'$ the particle is now located at a site with distance $\vec{D}$ from the starting point (we assume the sublattice is the same). The time-evolution operator looks like this:
\begin{equation}
    U(\mathbf{k},t')= \left( {\begin{array}{cc}
    e^{i\phi_0}e^{i\mathbf{k}\cdot\vec{D}} & 0 \\
    0 & e^{i\phi_1}e^{-i\mathbf{k}\cdot\vec{D}} \\
  \end{array} } \right)
\end{equation}
$\phi_{1,0}$ are phases picked up during the evolution which do not influence the center of mass position, given only by $i\bra{0}{U}^{-1}\nabla {U}\ket{0}=ie^{-i\phi_0}e^{-i\mathbf{k}\cdot\vec{D}}\nabla (e^{-i\phi_0}e^{i\mathbf{k}\cdot\vec{D}})=-\vec{D}$.

From that point, tunneling is turned on along just one precise direction to another lattice site on the other sublattice. Then
this couples the two sublattices with a fixed amplitude but with a quasimomentum-dependent phase:
$H_{01}(\mathbf{k},t)\propto e^{i\mathbf{k}\cdot \vec{a}}$ where $\vec{a}$ is the vector connecting the two lattice sites involved in the tunneling.

${U}$ in the time interval $[t',~t'+t_{\mathrm{flop}}]$  can be written as:
\begin{equation}    
\begin{aligned}    {U}(\mathbf{k},t)=& \left( {\begin{array}{cc}
    \mathrm{cos}(\theta)e^{i\phi_0}e^{i\mathbf{k}\cdot\vec{D}} & \mathrm{sin}(\theta)e^{i\phi_{10}}e^{i\mathbf{k}\cdot(\vec{D}+\vec{a})} \\
    \mathrm{sin}(\theta)e^{i\phi_{01}}e^{-i\mathbf{k}\cdot(\vec{D}+\vec{a})} & \mathrm{cos}(\theta)e^{i\phi_1}e^{-i\mathbf{k}\cdot\vec{D}} \\
  \end{array} } \right)\\
    {U}^{-1}(\mathbf{k},t)=& \left( {\begin{array}{cc}
    \mathrm{cos}(\theta)e^{-i\phi_0}e^{-i\mathbf{k}\cdot\vec{D}} & \mathrm{sin}(\theta)e^{-i\phi_{01}}e^{i\mathbf{k}\cdot(\vec{D}+\vec{a})} \\
    \mathrm{sin}(\theta)e^{-i\phi_{10}}e^{-i\mathbf{k}\cdot(\vec{D}+\vec{a})} & \mathrm{cos}(\theta)e^{-i\phi_1}e^{i\mathbf{k}\cdot\vec{D}} \\
  \end{array} } \right)\end{aligned}\label{flop}
\end{equation}
where $\theta,~\phi$ are time-dependent but not quasi-momentum dependent. In particular, $\theta(t')=0,~\theta(t'+t_\mathrm{flop}=\pi/2)$. Note that the case of a tunneling where the start- and target sublattices are inverted is obtained by taking $\theta(t')=\pi/2,~, \theta(t'+t_\mathrm{flop}=0)$, also the description we make now covers all cases.

We demonstrate that  $\bra{s}\delta B\ket{s}\xrightarrow{}0$.
For that we need to calculate $\partial_t{U}^{-1}$:\begin{equation}\begin{aligned}
    &\partial_t{U}^{-1}(\mathbf{k},t'+t)=\\
    & \Dot{\theta}\left( {\begin{array}{cc}
    -\mathrm{sin}(\theta)e^{-i\phi_0}e^{-i\mathbf{k}\cdot\vec{D}} & \mathrm{cos}(\theta)e^{-i\phi_{01}}e^{i\mathbf{k}\cdot(\vec{D}+\vec{a})} \\
    \mathrm{cos}(\theta)e^{-i\phi_{10}}e^{-i\mathbf{k}\cdot(\vec{D}+\vec{a})} & -\mathrm{sin}(\theta)e^{-i\phi_1}e^{i\mathbf{k}\cdot\vec{D}}\\
  \end{array} } \right)+\\
  &i \left( {\begin{array}{cc}
    -\Dot{\phi}_0\mathrm{cos}(\theta)e^{-i\phi_0}e^{-i\mathbf{k}\cdot\vec{D}} & -\Dot{\phi}_{01}\mathrm{sin}(\theta)e^{-i\phi_{01}}e^{i\mathbf{k}\cdot(\vec{D}+\vec{a})} \\
    -\Dot{\phi}_{10}\mathrm{sin}(\theta)e^{-i\phi_{10}}e^{-i\mathbf{k}\cdot(\vec{D}+\vec{a})} & -\Dot{\phi}_1\mathrm{cos}(\theta)e^{-i\phi_1}e^{i\mathbf{k}\cdot\vec{D}} \\
  \end{array} } \right)
  \end{aligned}
\end{equation}

We calculate $\bra{0}B\ket{0}$ in this limit:
\begin{equation}
\begin{aligned}&2i\bra{0}\partial_t{U}^{-1}\nabla {U}\ket{0}=\\&2i\bra{0}\partial_t{U}^{-1}\ket{0}\bra{0}\nabla {U}\ket{0}+i\bra{0}{U}^{-1}\ket{1}\bra{1}\nabla {U}\ket{0}=
\\
&-2i\Dot{\theta}\;\mathrm{sin}(\theta)e^{-i\phi_0}e^{-ik\cdot\vec{D}}\nabla (\mathrm{cos}(\theta)e^{-i\phi_0}e^{ik\cdot\vec{D}})+\\
&2i\Dot{\theta}\;\mathrm{cos}(\theta)e^{i\phi_{10}}e^{-ik\cdot\vec{D}}\nabla (\mathrm{sin}(\theta)e^{-i\phi_{10}}e^{ik\cdot(\vec{D}+\vec{a})})=\\&-2\Dot{\theta}\;\mathrm{sin}(\theta)\mathrm{cos}(\theta)\;\vec{a}=-\Dot{\theta}\;\mathrm{sin}(2\theta)\;\vec{a}
\end{aligned}\end{equation}

We neglected the terms proportional to the derivatives of $\phi_0,~\phi_1,~\phi_{10},~\phi_{01}$ because they give rise to a purely complex contribution to the velocity. 
Similarly, one obtains for $\bra{0}\delta C\ket{0}$:
\begin{equation}
\begin{aligned}&i\bra{0}{U}^{-1}\nabla {U}\ket{0}=\\&i\bra{0}{U}^{-1}\ket{0}\bra{0}\nabla {U}\ket{0}+i\bra{0}{U}^{-1}\ket{1}\bra{1}\nabla {U}\ket{0}=
\\
&i\mathrm{cos}(\theta)e^{-i\phi_0}e^{-i\mathbf{}\cdot\vec{D}}\nabla (\mathrm{cos}(\theta)e^{-i\phi_0}e^{i\mathbf{}\cdot\vec{D}})+\\
&i\mathrm{sin}(\theta)e^{i\phi_{10}}e^{-i\mathbf{}\cdot\vec{D}}\nabla (\mathrm{sin}(\theta)e^{-i\phi_{10}}e^{i\mathbf{}\cdot(\vec{D}+\vec{a})})=\\&-\mathrm{cos}^2(\theta)\vec{D}-\mathrm{sin}^2(\theta)(\vec{D}+\vec{a})
\end{aligned}\end{equation}

Note that $\bra{1}C\ket{1}=-\bra{0}C\ket{0},~\bra{1}B\ket{1}=-\bra{0}B\ket{0}$.
Because there is no dependence on quasimomentum,  $\bra{s}\delta B\ket{s},\bra{s}\delta C\ket{s}\xrightarrow{}0$, hence the position-velocity correlation term tends to zero as well. 

We now consider the band-flattening term. There are now two cases to be considered: whether the wavepacket comes back to the initial lattice site \cite{Rudner2013}, or whether it lands on another site on the opposite sublattice \cite{Kitagawa2010}. (If the particle ends up on a different site on the same sublattice as the starting one, one has net transport, and hence we are not dealing with an anomalous phase.) In the first case:
\begin{equation}
    {U}(T,k)= \left( {\begin{array}{cc}
    e^{i\phi_0} & 0\\
    0 & e^{i\phi_1} \\
  \end{array} } \right)
\end{equation}
with no dependence on $k$. Therefore also $H_F$ is $k-$independent and hence $\int_T^{2T}B\times C=0$ because of the derivatives w.r.t. $k$ there present.
In the second case:
\begin{equation}
    U(T)=\left( {\begin{array}{cc}
    0 & e^{i\phi_{10}}e^{i\mathbf{k}\cdot(\vec{D}+\vec{a})} \\
    e^{i\phi_{01}}e^{-i\mathbf{k}\cdot(\vec{D}+\vec{a})} & 0 \\
  \end{array} } \right)
\end{equation}
and hence the time evolution in the $[~T,2T~]$ interval is described similarly as in Eq.~(\ref{flop}), with a transfer from lattice site $\vec{D}+\Vec{a}$ (or -$\vec{D}-\Vec{a}$) back to $0$. There we can decompose $\overline{B\times C}$ into $\overline{B}\times \overline{C}$ because, as before, the deviations from the mean are zero. In a single transfer one has $B\parallel C$ and hence this extra term is also here zero. 

As an alternative way to study this second case, one could define two physical driving periods as the Floquet period and, because now the localized particle comes back to the original position, same conclusions apply as in the first case. (Note that in the doubled period, both the area and the winding number would be doubled.)

\end{document}